\documentclass[12pt,preprint]{aastex}

\def\kms{\ifmmode{\rm km\th s^{-1}}\else km\th s$^{-1}$\fi}
\def\th{\thinspace}

\shortauthors{Boden et al.}
\shorttitle{V773~Tau~A}

\begin{document}

\title{Dynamical Masses for Pre-Main Sequence Stars: A
Preliminary Physical Orbit for V773 Tau A}

\email{**Accepted Form -- 15 June 2007**}

\author{Andrew F.\ Boden\altaffilmark{1,2},
        Guillermo Torres\altaffilmark{4},
        Anneila I.\ Sargent\altaffilmark{3},
        Rachel L.\ Akeson\altaffilmark{1},
        John M.\ Carpenter\altaffilmark{3},
	David A.\ Boboltz\altaffilmark{5},
	Maria Massi\altaffilmark{6},
	Andrea M.~Ghez\altaffilmark{7},
	David W.\ Latham\altaffilmark{4},
	Kenneth J.~Johnston\altaffilmark{5},
	Karl M.~Menten\altaffilmark{6},
	Eduardo Ros\altaffilmark{6}
}

\email{bode@ipac.caltech.edu}

\altaffiltext{1}{Michelson Science Center, California
Institute of Technology, 770 South Wilson Ave., Pasadena CA 91125}

\altaffiltext{2}{Department of Physics and Astronomy, Georgia State
University, 29 Peachtree Center Ave., Science Annex, Suite 400,
Atlanta GA 30303}

\altaffiltext{3}{Division of Physics, Math, and Astronomy, California
Institute of Technology, MS 105-24, Pasadena, CA 91125}

\altaffiltext{4}{Harvard-Smithsonian Center for Astrophysics, 60
Garden St., Cambridge MA 02138}

\altaffiltext{5}{U.S.~Naval Observatory, 3450 Massachusetts Ave, Washington DC 20392}

\altaffiltext{6}{Max-Planck-Institut f\"ur Radioastronomie, Auf dem H\"ugel 69, D-53121 Bonn, Germany}

\altaffiltext{7}{Division of Astronomy and Astrophysics, UCLA, Los Angeles, CA 90095-1562}


\begin{abstract}

We report on interferometric and radial-velocity observations of the
double-lined 51-d period binary (A) component of the quadruple
pre-main sequence (PMS) system V773 Tau.  With these observations we
have estimated preliminary visual and physical orbits of the
V773~Tau~A subsystem.  Among other parameters, our orbit model
includes an inclination of 66.0 $\pm$ 2.4 deg, and allows us to infer
the component dynamical masses and system distance.  In particular we
find component masses of 1.54 $\pm$ 0.14 and 1.332 $\pm$ 0.097
M$_{\sun}$ for the Aa (primary) and Ab (secondary) components
respectively.

Our modeling of the subsystem component spectral energy distributions
finds temperatures and luminosities consistent with previous studies,
and coupled with the component mass estimates allows for comparison
with PMS stellar models in the intermediate-mass range.  We compare
V773~Tau~A component properties with several popular solar-composition
models for intermediate-mass PMS stars.  All models predict 
masses consistent to within 2-sigma of the dynamically determined 
values, though some models predict values that are more consistent than 
others.

\end{abstract}


\section{Introduction}
\label{sec:introduction}

Accurate determinations of the physical properties of stars
(e.g.~mass, radius, temperature, luminosity, elemental abundances,
rotation, etc.) provide fundamental tests of stellar structure and
evolution models.  Among the areas where our understanding of stellar
structure is most uncertain is in pre-main sequence (PMS) stars,
particularly for low-mass systems \citep[see][and references for
summaries]{Palla2001,Hillenbrand2004,Mathieu2007}.  Previously we have
reported results from our program of studying PMS binary systems with
the intent of establishing dynamical mass and luminosity constraints
\citep{Boden2005b}.  Here we report on observations of the double-lined
spectroscopic binary V773~Tau~A, part of the PMS quadruple system
V773~Tau.

\objectname[V773 Tau]{V773~Tau} (HDE~283447, HBC~367) is among the
most interesting and well-studied systems in the Taurus-Auriga star
forming region.  V773~Tau was first identified as a T~Tauri star by
\citet{Rydgren1976} based on H$\alpha$ and Ca II H and K emission,
high lithium abundance, photometric variability, and K2 spectral type.
The object presented an enigmatic mixture of classical (CTTS) and
weak-lined T Tauri (WTTS) characteristics until it became apparent
there are multiple components; V773~Tau was resolved as a visual
binary independently by \citet{Ghez1993} and \citet{Leinert1993} (with
components designated here as A and B).  \citet{Martin1994} suggested,
and \citet[herein W1995]{Welty1995} established A as a short-period
(51 d) double-lined spectroscopic binary.  In high-angular resolution
studies \citet[herein D2003]{Duchene2003} and \citet{Woitas2003}
independently identified an additional, ``infrared'' component in the
system (herein designated C -- note D2003 use an alternate component
notation), making V773~Tau an apparent compact quadruple system with
four stars within roughly 100 AU \footnote{There are at least three
different component notations used in describing the V773~Tau system.
Herein we have described the 51-d SB2 subsystem as ``A'', with stellar
components Aa and Ab; this is consistent with nomenclature used in
\citet{Massi2006} and \citet{Woitas2003}.  Conversely, D2003 refer to
the 51-d SB2 subsystem as ``AB''.  Finally, based on arguments made by
\citet{Hartigan1994} on the ``likely'' association of V773~Tau and
\objectname[FM Tau]{FM~Tau}, the Washington Double Star Catalog
\citep{Mason2001} describes the four-star complex as ``A'' (with
FM~Tau as ``B''), and the 51-d SB2 as ``Aa'' with components ``Aa1''
and ``Aa2''.}.  Using spatially-resolved photometry, D2003 find the A,
B, and C components to have distinctly different spectral energy
distributions.  D2003 conclude that V773~Tau contains stars at very
different PMS evolutionary states: the optically brighter A stars are
WTTS in character, while the circumstellar material inferred from the
significant IR excess is associated with the B and C components.

Further, the V773~Tau~A subsystem is a bright, polarized, and
highly-variable radio source.  \citet{Kutner1986} discovered radio
continuum emission from the system, appearing as the most luminous
Taurus-Auriga T Tauri star in the survey by \citet{O'Neal1990}.
\citet{Feigelson1994} published a comprehensive multiwavelength study
of the V773~Tau system, observing both polarized and flaring radio
emission.  W1995 argued that both A components are chromospherically
active based on variations in their optical spectrum and estimates of
relatively rapid rotation ($v \sin i$ $\sim$ 40 km s$^{-1}$); this
point was confirmed by \citet[herein P1996]{Phillips1996}, who
resolved the radio emission into a clear binary morphology apparently
corresponding to the two A components.  \citet{Massi2002} identified
an increase in flare activity around A periastron, and
\citet{Massi2006} studied one such flare in detail; more frequent and
powerful radio flaring at periastron is apparently due to interaction
of component magnetic fields while in close proximity.  Finally,
\citet[herein L1999]{Lestrade99} used Very Long Baseline
Interferometry (VLBI) astrometry to measure the trigonometric parallax
for V773~Tau~A as 6.74 $\pm$ 0.25 mas, corresponding to a system
distance of 148.4 $\pm$ 5.5 pc.

Here we report on observations of the V773~Tau~A subsystem made with
the Keck Interferometer \citep[KI, see][]{Colavita2003}, the Very Long
Baseline Array \citep[VLBA, see][]{Napier1994,Zensus1995}, and
spectroscopic radial velocities from the Harvard-Smithsonian Center
for Astrophysics (CfA) ``Digital Speedometers'' \citep{Latham1992}.
These observations resolve V773~Tau~A and allow us to estimate the
visual and physical (i.e.~three-dimensional) orbits, and determine the
component dynamical masses and system distance.  We describe these
orbit models as {\em preliminary} because they weakly depend on
ongoing observations and modeling of the larger V773~Tau system.
Combining derived dynamical masses and system distance with component
radiometric modeling, we estimate component temperatures and
luminosities, allowing us to make direct comparisons with a number of
popular solar-composition PMS stellar models.

\section{Observations and Orbital Solution}
\label{sec:observations}

\paragraph{KI Observations}
The KI interferometric observable used for these measurements is the
fringe contrast or {\em visibility} (squared, V$^2$) of an observed
brightness distribution on the sky.  Orbital analysis methods for such
wide-band V$^2$ observations are discussed in \citet{Boden2000} and
not repeated here.  V773~Tau~A was observed by KI in $K$-band
($\lambda \sim 2.2 \mu$m) on five nights between 2002 Oct 24 and 2006
Dec 8, a data set spanning roughly four years and 30 orbital periods.
V773~Tau~A and calibration objects were typically observed multiple
times during each of these nights, and each observation (scan) was
approximately 130 sec long.  For each interferometric scan we computed
a mean \citep[incoherent wide-band;][]{Colavita1999,Colavita2003}
V$^2$ value from the scan data, and the error in the V$^2$ estimate is
inferred from the rms internal scatter.  For this analysis we have
used \objectname[HD 27741]{HD~27741} (G0 V) and \objectname[HD
28483]{HD~28483} (F6 V) as calibration objects.  Calibrating our
interferometric data with respect to these objects results in 27
calibrated visibility scans on V773~Tau~A over the five epochs.
Because the AO-corrected Keck Telescopes separately resolve the
V773~Tau A-B-C complex (D2003), and the KI beam combiner is fed by
single-mode fiber, no significant light from the B or C components
enters the fringe camera while observing A \citep[the $K$-band
projected FWHM size of the KI single-mode fiber is 50 mas, so the B
component is approximately five e-folding lengths away from A at the
epoch of our observations;][D2003]{Colavita2003}. Consequently no
special provisions for extra light are necessary in processing KI
observations of V773~Tau~A.

\begin{table}
\dummytable\label{tab:V2Table}
\end{table}

\begin{figure}[ph]
\epsscale{0.9}
\plotone{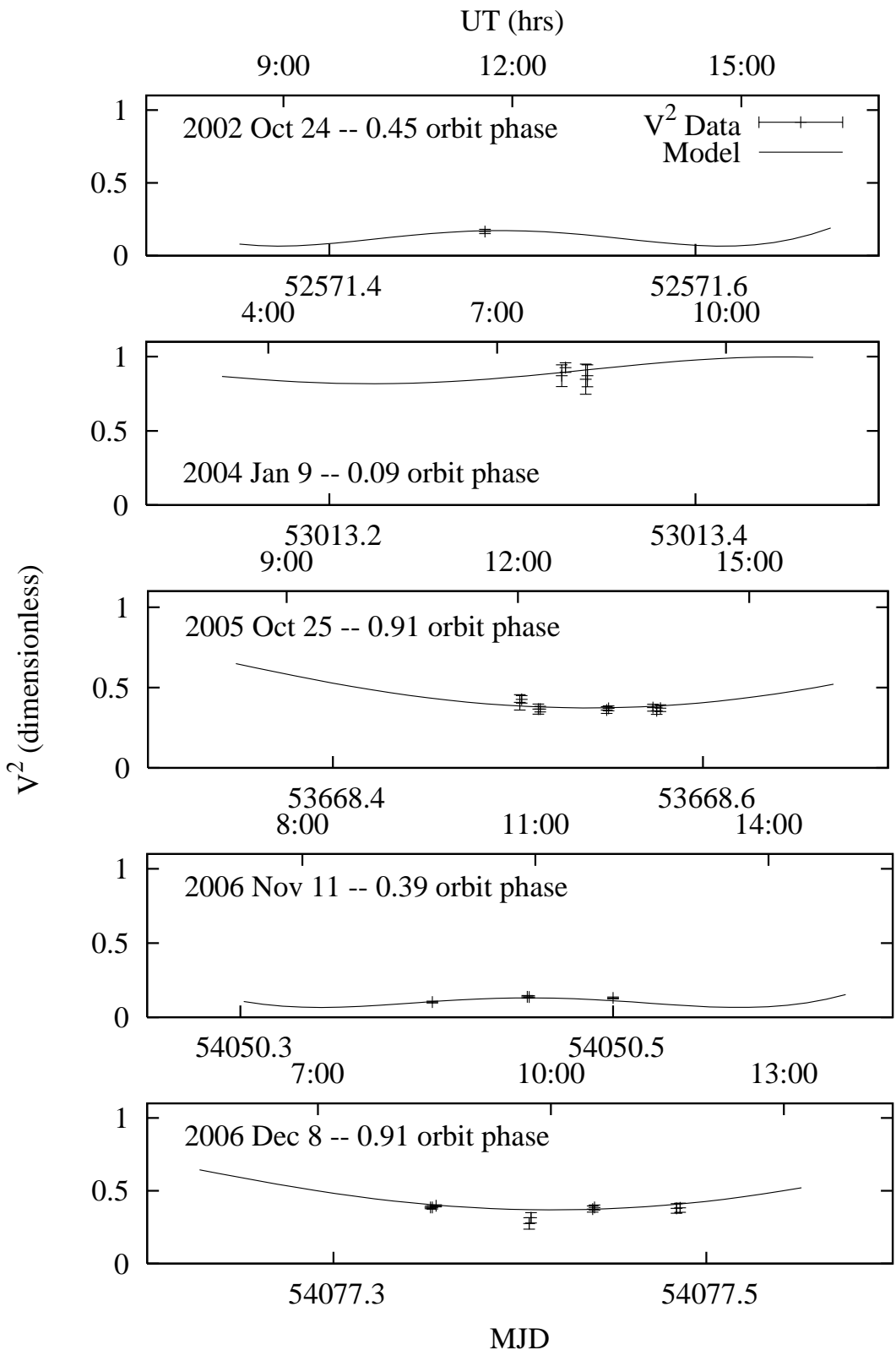}
\caption{KI Data/Model Comparisons for V773~Tau~A V$^2$ Observations.
Here we give comparisons for our five epochs of KI V$^2$ observations
and V$^2$ predictions from our ``Hybrid'' orbital model
(Table~\ref{tab:orbit}).  In each case the data and model are shown.
\label{fig:v2Data}}
\end{figure}

Our KI V$^2$ observations are summarized in Table \ref{tab:V2Table}.
The V$^2$ data set is depicted in Figure~\ref{fig:v2Data}, along with
V$^2$ predictions from our ``Hybrid'' orbit model (discussed below).

\paragraph{VLBI Observations}
P1996 observed V773~Tau~A with the VLBA, and with similar spatial
frequencies as our KI observations ($\sim$ 50 M$\lambda$) resolved the
typical 2 mas separation between the A components.  P1996 Figure 6
depicts a VLBA contour map of V773~Tau~A at a particular epoch (1992
Sept 11), and clearly shows a binary source structure.  P1996
interpreted this binary morphology as radio flare emission from the
two subsystem components.  P1996 argued that the radio emission is
coincident with the stars (to within a ``few'' stellar diameters), so
a component separation interpreted as at-epoch relative astrometry
was estimated by P1996.

\begin{figure}[ph]
\epsscale{0.6}
\plotone{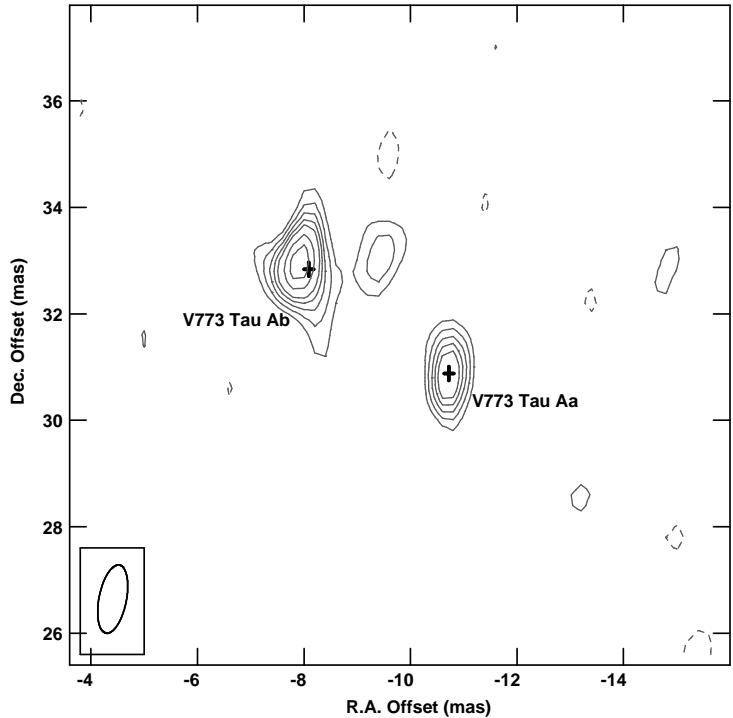}
\caption{Sample VLBA Map from 2004 March 15.  Here we give a contour
map of the total intensity continuum emission toward V773~Tau~A from
VLBA+EB observations of 2004 March 15.  Contours are at -3, 3, 4.5, 6,
7.5, 9, 12, and 15 $\sigma$ of the rms noise in the image (0.1
mJy/beam).  The FWHM beam size (lower left) is 1.36$\times $0.50~mas
at a position angle of $-$11.6$^{\circ}$.  The peak flux density in
the image is 0.55~mJy/beam.  V773~Tau~A is offset in these data and
image as a calibrator (J0403+2600, not shown) served as the phase
center.  Heavy crosses are rendered at the predicted stellar
separation based on our ``V$^2$ \& RV'' orbit from
Table~\ref{tab:orbit} (i.e. no radio data included in the orbit
modeling), registered to the center of the apparent primary position.
The sizes of the crosses indicate the uncertainty in the separation
estimate ($\sim$ 0.1 mas per axis).
\label{fig:VLBAmap}}
\end{figure}

\begin{figure}[p]
\epsscale{0.8}
\plotone{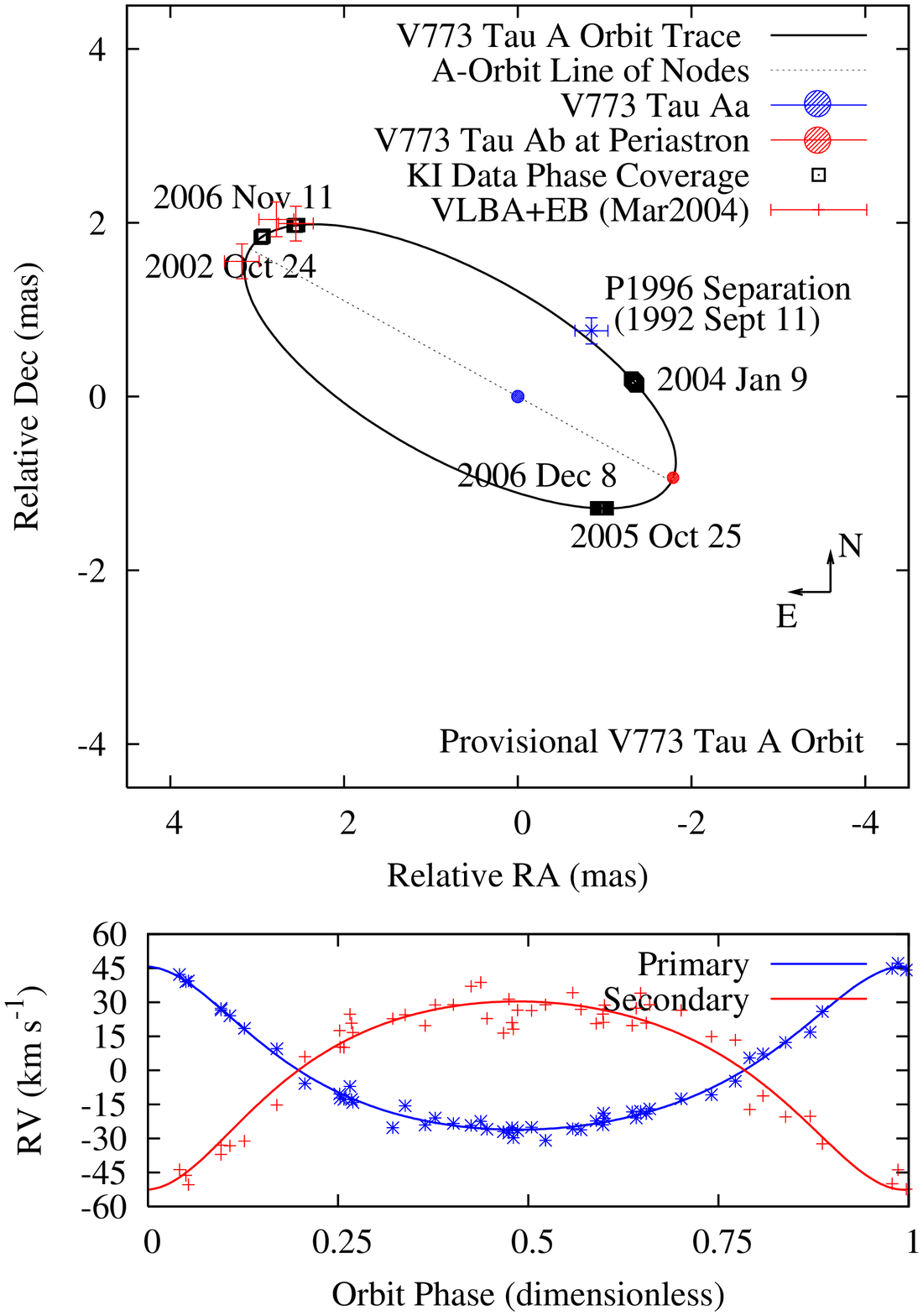}
\caption{Orbit of V773~Tau~A as derived in our ``Hybrid'' solution
(Table~\ref{tab:orbit}).  Upper Panel: the relative visual orbit model
of V773~Tau~A is shown, with the primary and secondary objects
rendered at T$_0$ (periastron).  The specific epochs where we have KI
V$^2$ phase coverage are indicated on the relative orbit (they are not
separation vector estimates), and the VLBI separations from P1996 (in
blue) and this work (in red) are also shown.  Component diameter
values are estimated and rendered to scale.  Lower Panel: the
double-lined radial velocity orbit model and data described here.
\label{fig:orbitPlot}}
\end{figure}

A team led by one of us (M.M.) has studied the physics of V773~Tau~A
radio flares.  In addition to previous results
\citep[e.g.][]{Massi2002,Massi2006}, V773~Tau~A was observed using the
VLBA in coordination with the Effelsberg 100\,m radio telescope
(herein VLBA+EB).  These 8.4-GHz (X-band) observations were conducted
around apastron on seven consecutive days (2004 March 11-17
inclusive).  As in the case of P1996, analysis of these observations
revealed what appears to be both A components simultaneously
exhibiting radio flares in most of these observations.  However some
of these ``double-flare'' observations show complex emission
morphologies, complicating estimation of the binary relative
separation.  Reviewing the maps from these seven epochs, we decided to
use astrometry from observations where there was significant
(i.e.~that the peak emission was greater than six sigma above the rms
noise level) emission from both components, and the component
structures were largely unresolved compared to the VLBA synthetic
beam.  These criteria resulted in our retaining three of the seven
2004 epochs for orbital analysis purposes; the astrometry resulting
from these observations is summarized in
Table~\ref{tab:VLBIastrometry}.  A sample image constructed from one
of the three astrometric epochs (2004 March 15) is shown in
Figure~\ref{fig:VLBAmap}.  The astrometric results of both the P1996
and these VLBI observations are depicted in
Figure~\ref{fig:orbitPlot}; these observations agree well with our
orbit model for V773~Tau~A described below.  The radio morphologies
seen in the full VLBI+EB data set will analyzed in a forthcoming
publication (Massi et al 2007, in prep).

\begin{table}
\dummytable\label{tab:VLBIastrometry}
\end{table}

\paragraph{RV Observations}
W1995 established V773~Tau~A as a double-lined spectroscopic binary,
and published an orbital solution.  RV monitoring of V773~Tau with the
CfA ``Digital Speedometers'' \citep{Latham1992} was carried out at the
Oak Ridge Observatory (Harvard, Massachusetts), the F.L.~Whipple
Observatory (Mount Hopkins, Arizona), and the MMT Observatory (also on
Mount Hopkins, Arizona).  The nearly identical instruments at these
facilities record a single echelle order and provide a wavelength
coverage of 45~\AA\ centered at 5187~\AA, with a resolving power of
$\lambda/\Delta\lambda \approx 35,\!000$. Wavelength calibration was
achieved in the usual manner through exposures of a Th-Ar lamp taken
before and after each science exposure.  The velocity zero point was
monitored by means of exposures of the dusk and dawn sky, and small
run-to-run corrections were applied to the velocities derived below as
described by \cite{Stefanik:99}.

Our spectroscopic observations of V773~Tau began roughly at the same
time as the W1995 data set and have continued to the present. In total
we have collected 52 spectra spanning the period 1991 Sep 25 to 2007
Jan 7, with signal-to-noise ratios of 16--35 per resolution element of
8.5~\kms.  The Radial Velocities (RVs) derived from these observations
are summarized in Table~\ref{tab:RVtable}.  RVs were derived using the
two-dimensional cross-correlation technique TODCOR \citep{Zucker:94},
with templates adopted from a large library of calculated spectra
based on Kurucz model atmospheres \citep[see also][]{Nordstrom:94,
Latham:02}. The best-matching templates were chosen by running large
grids of cross-correlations as described by \citet[][]{Torres2003},
which also yielded estimates of some of the stellar parameters. For an
assumed Taurus solar metallicity \citep[see][]{Padgett1996} and
surface gravities appropriate for these stars ($\log g \approx 4.0$;
see Table~5), we derived effective temperatures of $4900 \pm 150$~K
and $4740 \pm 200$~K for Aa and Ab, respectively.  The adopted errors
are conservative estimates to account for the abundance
uncertainty. Formal errors are 50~K smaller, and are based on the
scatter from the different spectra. Using the same methods we
estimated the projected rotational velocity for the primary star to be
$v \sin i = 38 \pm 4$~\kms, in good agreement with the determination
by W1995 (41.4~\kms; no uncertainty reported). We were not able to
determine the $v \sin i$ for the fainter secondary, possibly due to
the relatively low signal-to-noise ratios of our spectra.  The
templates for our radial-velocity determinations were selected to have
temperatures of 5000~K and 4750~K (the nearest values in our template
library) for Aa and Ab, and rotational velocities of 40~\kms\ for both
stars. The light ratio between the secondary and the primary was
determined with TODCOR by leaving it as an additional free parameter
during the velocity determinations, as described by \cite{Zucker:94}.
We obtained $\ell_{Ab}$/$\ell_{Aa}$ = 0.37 $\pm$ 0.03 at a mean
wavelength of 5187~\AA.

The presence of two additional spatially unresolved components in the
V773~Tau system may in principle affect both our determination of the
stellar parameters of Aa and Ab and their radial velocities. While the
infrared companion V773~Tau~C \citep[D2003,][]{Woitas2003} is much too
faint to cause any contamination, V773~Tau~B is estimated to be
$\sim$1.7 magnitudes fainter than the spectroscopic binary in the
optical (D2003), and could possibly have an effect.  We attempted to
detect the lines of this component in our spectra using an extension
of TODCOR to three dimensions \citep[see][]{Zucker:95}, but were
unsuccessful. On this basis we conclude that it does not significantly
affect our results above. However, its presence does complicate the
analysis of the RV data for the spectroscopic binary in that the
motion in the A-B orbit needs to be taken into account. We discuss
this further below, and in a forthcoming publication (Torres et al
2007, in prep).

\begin{table}
\dummytable\label{tab:RVtable}
\end{table}

\paragraph{Orbital Solution}
We have combined the astrometric (KI, P1996, VLBA+EB) and radial
velocity data described above to estimate the visual and physical
(i.e.~three-dimensional) orbits for V773~Tau~A.  In particular,
methods for modeling the wide-band visibility data used in this
analysis are described in \citet{Boden2000}, and not repeated here.
Our ``Hybrid'' orbital solution (Table~\ref{tab:orbit}) is depicted in
Figure~\ref{fig:orbitPlot}.  The upper panel depicts the relative
visual orbit model, with the primary (Aa) component rendered at the
origin, and the secondary (Ab) component rendered at periastron.  The
KI V$^2$ data phase coverage is indicated on the visual orbit with
points (they are {\em not} separation vectors); the phase coverage of
the V$^2$ data is sparse relative to other similar analyses
\citep[e.g.][]{Boden2000}, but phase coverage provided by VLBI
observations from P1996 and this work complements the V$^2$ coverage.
Further, the incorporation of the VLBI separations also breaks the
180$^\circ$-$\Omega$ degeneracy inherent in $V^2$-only analyses
\citep[e.g.][]{Boden2000,Boden2005a}, so the A-orbit rendered in
Figure~\ref{fig:orbitPlot} is indeed as it appears on the sky.  The
apparent size of the V773~Tau~A components are estimated
(\S~\ref{sec:SED}, Table~\ref{tab:physics}) and rendered to scale;
these same diameter values are used in the orbital modeling.  The
lower panel depicts radial velocity curves from the ``Hybrid'' orbit
model and radial velocity data described above.

V773~Tau~A presents two significant challenges for RV orbit
estimation.  The first is that both components are relatively rapid
rotators: W1995 found the $v \sin i$ for both stars to be around 40 km
s$^{-1}$; this rotational broadening is seen in our spectra as well,
and it significantly degrades the precision of the velocities.
Secondly, and most notably, is the A-barycenter motion that results
from the A-B orbit.  D2003 Figure 4 depicts the significant A-B
orbital motion over the time since the detection of V773 Tau B, and
our RV data set (starting in 1991) not only spans essentially all of
this period, but includes the A-B periastron passage expected in 1996
(D2003).  It is therefore important to account for the A-B orbital
motion in the modeling of the A component RVs.  Modeling of the A-B
orbit would not have been possible at the time of W1995, but becomes
feasible with the present data set.

A separate RV-only orbital solution was carried out in which, in
addition to solving for preliminary elements of the A orbit, we
modeled the Keplerian motion of A around the A-B barycenter.  In this
analysis we held fixed some of the elements of the outer orbit to the
values given by D2003 (specifically, the period, eccentricity,
longitude of periastron, and time of periastron passage) and we solved
for the velocity semi-amplitude of the barycenter of A ($K_A = 6.11
\pm 0.34$ km s$^{-1}$).  This motion was then removed from our
original velocities prior to combining them with the KI and VLBI data.
A more complete treatment would incorporate all the astrometric
measurements for the A-B orbit (e.g., D2003) along with the velocities
and astrometry for the inner subsystem in a full solution solving for
both orbits simultaneously.  Such a study is beyond the scope of the
present work, but will be described in a forthcoming publication
(Torres et al 2007, in prep).

There are significant differences between several of our orbital
parameters for V773~Tau~A and the corresponding values from W1995,
most notably in the period and K$_{Aa}$ (see Table~\ref{tab:orbit}).
We attribute these differences to a combination of at least three
effects: {\it i}) limited phase coverage of the A orbit by W1995,
particularly near A apastron; {\it ii}) possible biases in the W1995
RVs due to line blending, exacerbated by the rapid rotation of Aa and
Ab along with the lower spectral resolution of that study compared to
ours; and {\it iii}) residual effects from the un-modeled motion in the
outer orbit in W1995's solution.

P1996 argue that their resolved VLBI image of V773~Tau~A measures the
relative component positions to within a ``few'' stellar diameters
(the primary apparent diameter is approximately 0.15 mas; see
\S~\ref{sec:SED}).  However, it is important to recognize that VLBI
and KI V$^2$ data measure different emission mechanisms, and further,
one would not expect individual radio flares to be centered on the
stellar photospheres (measured by the KI V$^2$ data).  This potential
photosphere/radio offset motivated us to approach integrating the
radio data into the orbit model cautiously.  In our initial screening
of the VLBI relative astrometry we compared the radio observations
with position predictions from a V$^2$ and RV orbit model
(Table~\ref{tab:orbit}).  We found the model reasonably reproduced the
observed radio separations with a typical residual of 0.2 -- 0.3 mas
($\approx$ 4 R$_*$), slightly larger than the formal errors expected
from the VLBI synthsized beam and the snr in the image, but in good
agreement with the expected IR/radio offsets.  The agreement we see
between the radio separations and the V$^2$ \& RV orbit model makes it
clear that the VLBI data do contain useful information on the
component separations.  We therefore felt confident in integrating the
KI V$^2$ and four VLBI observations (including the P1996 epoch) into
our orbit model estimation.  An analysis of the radio mophologies seen
in the VLBA+EB observations will be presented in a future publication
(Massi et al 2007, in prep).

Figure \ref{fig:v2Data} depicts direct comparisons between our KI
V$^2$ observations and predictions from our V773~Tau~A ``Hybrid''
orbit model from Table \ref{tab:orbit} (the five nights of data are
each rendered in separate sub-panels).  The model is seen to be in
good agreement with the KI data.  Further, Figure~\ref{fig:orbitPlot}
shows a direct comparison between the VLBA separations and the
``Hybrid'' visual orbit model.  Again, the body of the VLBI data are
in good agreement with the orbit model, and the VLBI phase coverage
complements the phase coverage provided by the KI V$^2$ data.

Orbital parameter estimates for the V773~Tau~A subsystem are
summarized in Table~\ref{tab:orbit}.  Included for relative comparison
are the orbital parameters determined by W1995, our RV double-lined
orbit (``RV''), the KI V$^2$ \& RV solution (``V$^2$ \& RV''), and the
full integrated solution including the RV, V$^2$, and VLBA data
(``Hybrid'').  The ``Hybrid'' designation for the composite orbit
model recognizes its unique inclusion of both near-IR and radio
interferometry data; this model is our favored orbit solution and is
used in the remaining analysis.  Again note that the A
barycenter velocity is artificially near zero in our orbital solutions
as the RV measurements have been corrected for our model of the A-B
orbital motion.

\begin{deluxetable}{lcccc}
\tablecolumns{5}
\tablewidth{0pc}

\tablecaption{V773~Tau~A Orbital Parameters.
\label{tab:orbit}
}

\tablecomments{Summarized here are
orbital parameters for the V773~Tau~A subsystem as determined by W1995
and present results.  We give three separate fits to our data sets: RV
only, KI V$^2$ integrated with RV (``V$^2$ \& RV''), and KI V$^2$,
VLBI, and RV (``Hybrid'') .  Barycenter velocity ($\gamma$) estimates
(shown in italics) are artificially near zero because we have
estimated and removed a model of the A-B orbital motion (see
discussion in text).  $\Omega$ is quoted in a position angle
convention.}

\tablehead{
\colhead{Orbital}        & \colhead{W1995}      & \multicolumn{3}{c}{This Work} \\
\colhead{Parameter}      &                      & \colhead{RV}             & \colhead{V$^2$ \& RV}  & \colhead{Hybrid}
}
\startdata
Period (d)               & 51.075 $\pm$ 0.018   & 51.1055 $\pm$ 0.0023     & 51.1033 $\pm$ 0.0022   &  51.1039  $\pm$ 0.0021 \\
T$_{0}$ (MJD)            & 49330.44 $\pm$ 0.56  & 53059.87 $\pm$ 0.36      & 53059.50 $\pm$ 0.35    &  53059.73 $\pm$ 0.33  \\
$e$                      & 0.267 $\pm$ 0.016    & 0.275 $\pm$ 0.012        & 0.2708 $\pm$ 0.0094    &  0.2717 $\pm$ 0.0085  \\
K$_{Aa}$ (km s$^{-1}$)   & 32.6 $\pm$ 0.7       & 35.95 $\pm$ 0.53         & 35.96  $\pm$ 0.53      &  35.90 $\pm$ 0.53     \\
K$_{Ab}$ (km s$^{-1}$)   & 43.1 $\pm$ 2.1       & 41.6 $\pm$ 1.4           & 41.5   $\pm$ 1.4       &  41.5 $\pm$ 1.4       \\
$\gamma$ (km s$^{-1}$)   & 13.8 $\pm$ 0.9       & {\em 0.00 $\pm$ 0.31}    & {\em 0.05  $\pm$ 0.31} &  {\em 0.02  $\pm$ 0.32} \\
$\omega_{Aa}$ (deg)      & 10.5 $\pm$ 5.1       & 5.1 $\pm$ 2.6            & 3.4   $\pm$ 2.5        &  4.6 $\pm$ 2.4        \\
$\Omega$ (deg)           &                      &                          & 61.5   $\pm$ 2.5       &  63.5 $\pm$ 1.7       \\
$i$ (deg)                &                      &                          & 65.9   $\pm$ 2.8       &  66.0  $\pm$ 2.4      \\
$a$ (mas)                &                      &                          & 2.777  $\pm$ 0.056     &  2.811  $\pm$ 0.047   \\
$\Delta$ $K$ (mag)       &                      &                          & 0.559  $\pm$ 0.082     &  0.551  $\pm$ 0.078  \\
\hline
\enddata

\end{deluxetable}

\section{V773~Tau~A physical properties}
\label{sec:physics}

The orbital parameters from Table~\ref{tab:orbit} allow us to directly
compute many of the physical properties of the V773~Tau~A subsystem
and its components.  Physical parameters derived from our ``Hybrid''
integrated visual/spectroscopic orbit are summarized in Table
\ref{tab:physics}.  The dynamical masses resulting from our orbit
model for the two A components are 1.54 $\pm$ 0.14 and 1.332 $\pm$
0.097 M$_{\sun}$ for Aa and Ab respectively.  This primary mass is in
acceptable agreement with inferences by \citet[herein
WG2001]{White2001} and \citet{Palla2001} based on radiometric
properties (i.e.~luminosity and effective temperaure).  Because it
bears on the A-B orbit analysis (Torres et al 2007, in prep) we
estimate the A subsystem mass as 2.87 $\pm$ 0.18 M$_{\sun}$.

L1999 estimated a V773~Tau system distance of 148.4 $\pm$ 5.5 pc.  The
distance determination to V773~Tau~A derived from our visual and
physical orbital solution is 136.2 $\pm$ 3.7 pc, corresponding to an
orbital parallax of 7.34 $\pm$ 0.20 mas, and consistent with the L1999
result at 8.1\% and 1.9-sigma.  L1999 describe their distance estimate
error as ``formal'', and it apparently did not account for the orbital
motion of A in their data reduction.  With these caveats we believe
the L1999 distance estimate and ours are in acceptable agreement.

At the distance of V773~Tau, neither of the A subsystem components are
significantly resolved by the KI K-band fringe spacing, and we must
resort to model diameters for the components.  We have estimated the
V773~Tau~A component apparent diameters through spectral energy
distribution modeling; details of the spectral energy distribution
modeling are given in \S\ref{sec:SED}.  We find apparent diameters of
0.152 $\pm$ 0.014 and 0.119 $\pm$ 0.011 mas for the Aa and Ab
components respectively.  With our system distance estimate these
estimated diameters correspond to physical radii of 2.22 $\pm$ 0.20
and 1.74 $\pm$ 0.19 R$_\sun$, and (combined with the mass estimates)
log surface gravities of 3.930 $\pm$ 0.094 and 4.081 $\pm$ 0.088 for
the Aa and Ab components respectively.  Similarly we have estimated
absolute magnitudes in $V$ and $K$, and a $V-K$ color index for each
of the components (Table~\ref{tab:physics}), but these are of limited
precision as they are dominated by the uncertainty in the extinction
to the system (\S~\ref{sec:SED}).

\begin{deluxetable}{ccc}
\tabletypesize{\small} \tablecolumns{3} \tablewidth{0pc}

\tablecaption{V773~Tau~A Physical Parameters.  
\label{tab:physics}
}

\tablecomments{Summarized here are
physical (and related) parameters for the V773~Tau~A subsystem as
derived primarily from the ``Hybrid'' orbit solution
(Table~\ref{tab:orbit}) and radiometric modeling (\S~\ref{sec:SED}).}

\tablehead{
\colhead{Physical}   & \colhead{Primary}    & \colhead{Secondary} \\
\colhead{Parameter}  & \colhead{(Aa)}       & \colhead{(Ab)} \\
}

\startdata
a (AU)               & 0.1777 $\pm$ 0.0042  & 0.2053 $\pm$ 0.0070  \\
Mass (M$_{\sun}$)    & 1.54  $\pm$ 0.14     & 1.332 $\pm$ 0.097  \\
\cline{2-3}
System Mass (M$_{\sun}$) & \multicolumn{2}{c}{2.87 $\pm$ 0.18} \\
System Distance (pc) & \multicolumn{2}{c}{136.2 $\pm$ 3.7} \\
$\pi_{orb}$ (mas)    & \multicolumn{2}{c}{7.34  $\pm$ 0.20} \\
\cline{2-3}
$T_{\rm eff}$ (K)    & 4900 $\pm$ 150       & 4740 $\pm$ 200   \\
Model Diameter (mas) & 0.152 $\pm$ 0.014    & 0.119 $\pm$ 0.011  \\
Bolometric Flux (10$^{-9}$erg cm$^{-2}$ s$^{-1}$)
                     & 4.44 $\pm$ 0.40   & 2.38 $\pm$ 0.15 \\
Luminosity (L$_{\sun}$)
                     & 2.56 $\pm$ 0.35   & 1.37 $\pm$ 0.15 \\
Radius (R$_{\sun}$)  & 2.22 $\pm$ 0.20   & 1.74 $\pm$ 0.19 \\
$\log g$             & 3.930 $\pm$ 0.094 & 4.081 $\pm$ 0.088 \\

M$_K$ (mag)          & 1.75  $\pm$ 0.12  & 2.30 $\pm$ 0.13   \\
M$_V$ (mag)          & 3.90  $\pm$ 0.16  & 4.98 $\pm$ 0.16   \\
$V$-$K$ (mag)        & 2.15  $\pm$ 0.14  & 2.68 $\pm$ 0.14   \\
\enddata

\end{deluxetable}

\subsection{Spectral Energy Distribution Modeling}
\label{sec:SED}
\begin{figure}[ph]
\includegraphics[angle=-90,width=14.5cm]{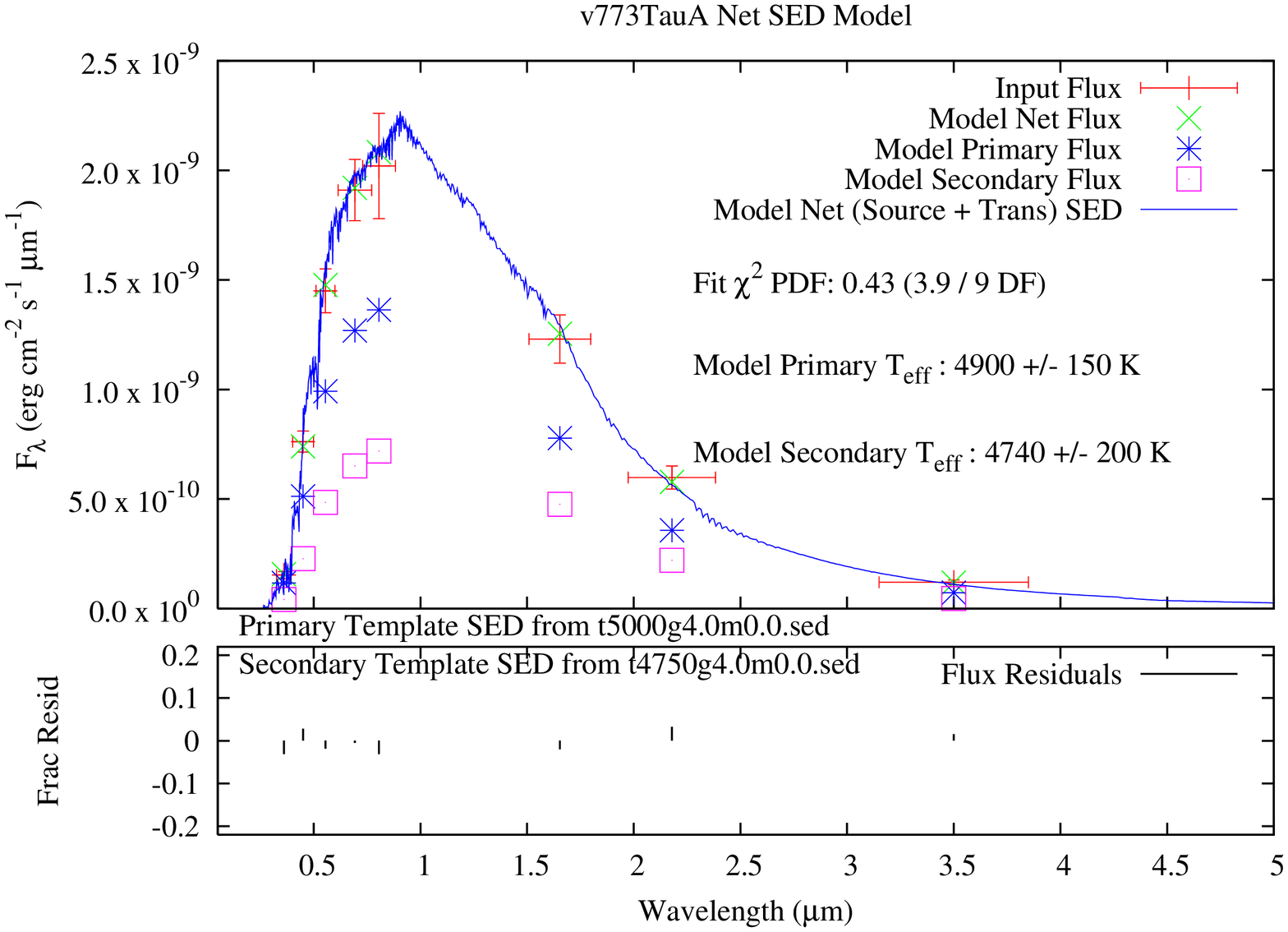}\\
\includegraphics[angle=-90,width=15.5cm]{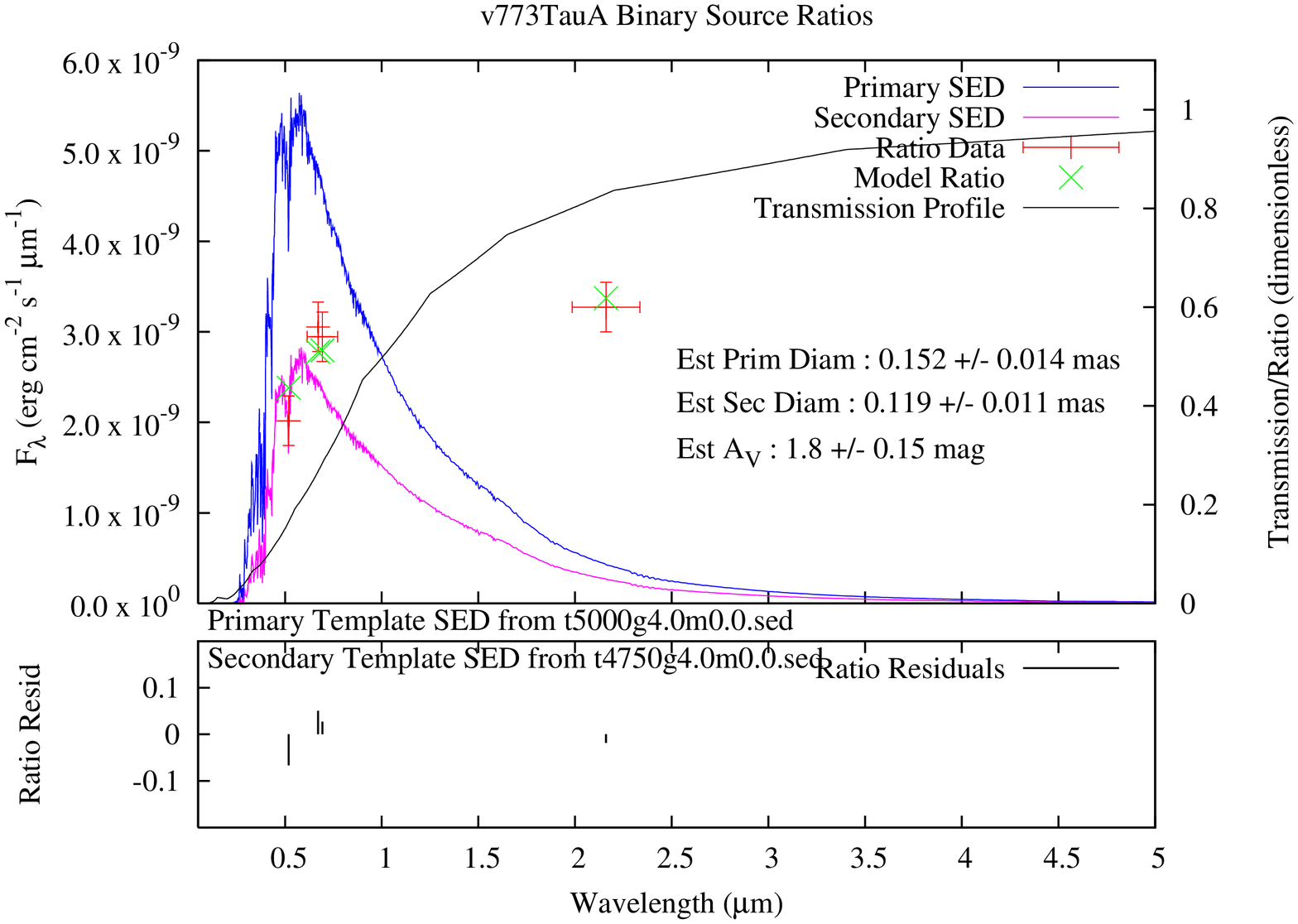}\\
\caption{Sample Spectral Energy Distribution Model for V773~Tau~A.
Here SED templates from \citet{Lejeune1997,Lejeune1998} have been used
to simultaneously model published flux measurements (from WG2001 \&
D2003) and flux ratio estimates (Table \ref{tab:orbit}; bottom) from
our KI V$^2$ and spectroscopic measurements.
\label{fig:V773TauA_SED}}
\end{figure}

Because interferometric observations potentially resolve the stellar
components in a binary system, we always construct spectral energy
distribution (SED) models for binary systems to estimate a priori
apparent diameters.  Sample results of our SED modeling of V773~Tau~A
are shown in Figure~\ref{fig:V773TauA_SED}.  Our modeling is based on
visible and near-IR photometry on A from WG2001 and D2003 respectively
(summarized in Table~\ref{tab:Fluxtable}), and component flux ratios
from W1995 and our visible spectroscopy and near-IR KI V$^2$ data
(discussed above).  Using a custom two-component SED modeling code, we
have modeled the A subsystem flux and ratios using solar-abundance,
intermediate surface gravity SED templates from \cite{KuruczModels}
and \citet{Lejeune1997,Lejeune1998}.  We model the flux and ratio data
with a large grid of SED templates for each component ranging in
effective temperature (spectral type) from 4250 -- 5500 K, surface
gravity of log g = 4.0, and including extinction as a free variable in
the parameter estimation.  Constrained by the ratio data, this SED
modeling preferred component temperature {\em differences} over a
reasonably narrow range ($\Delta$ $T_{\rm eff} \sim$ 150--250 K)
consistent with our spectroscopic study.  But because the component
colors are somewhat degenerate with reddening from dust extinction we
found acceptable solutions over a relatively broad range ($\sim$
500~K) of component temperatures.  We have therefore adopted templates
corresponding to the spectroscopic temperatures (4900 and 4740 K for
Aa and Ab respectively).  These temperatures acceptably agree with the
composite temperature estimated by WG2001, but the hotter primary in
our modeling prefers a slightly higher extinction value (A$_V$ = 1.80
$\pm$ 0.15 vs 1.39 $\pm$ 0.17 from WG2001).  The resulting component
bolometric fluxes and luminosities (the second based on our distance
determination) are summarized in Table~\ref{tab:physics}.  In
particular we find luminosities of 2.56 $\pm$ 0.35 and 1.37 $\pm$ 0.15
L$_\sun$ for Aa and Ab respectively; these values are in good
agreement with the range of previous component luminosity estimates
from W1995, \citet{Ghez1997}, and D2003.

\begin{table}
\dummytable\label{tab:Fluxtable}
\end{table}

\section{Discussion}
\label{sec:discussion}

\subsection{Comparison with stellar evolution models}
\label{sec:modelcomp}

\citet{Hillenbrand2004} and \citet{Mathieu2007} have recently
summarized comparisons between PMS stellar models and PMS stars with
dynamical mass determinations.  In particular \citet{Hillenbrand2004}
has shown that PMS models are in good agreement with
dynamically-determined masses above 1.2 M$_\sun$.  However, below 1.2
M$_\sun$ the existing models do a poorer job of matching observed
component properties, tending to predict hotter and more luminous
stars for a given mass.

\begin{figure}[tp]
\epsscale{1.1}
\plottwo{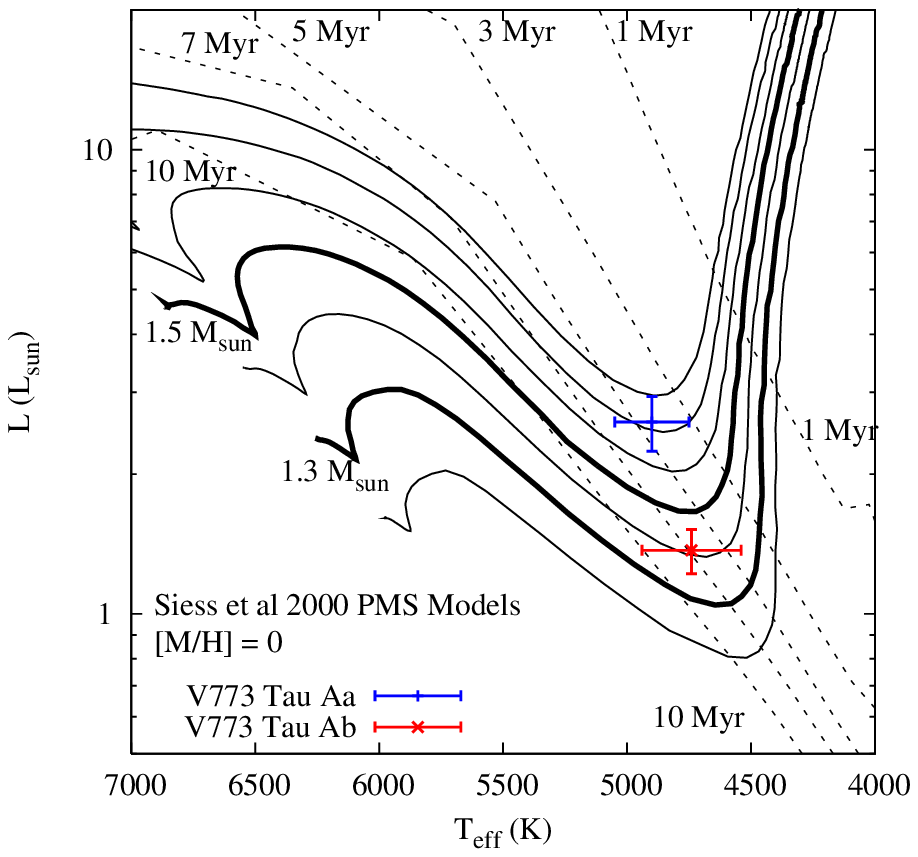}{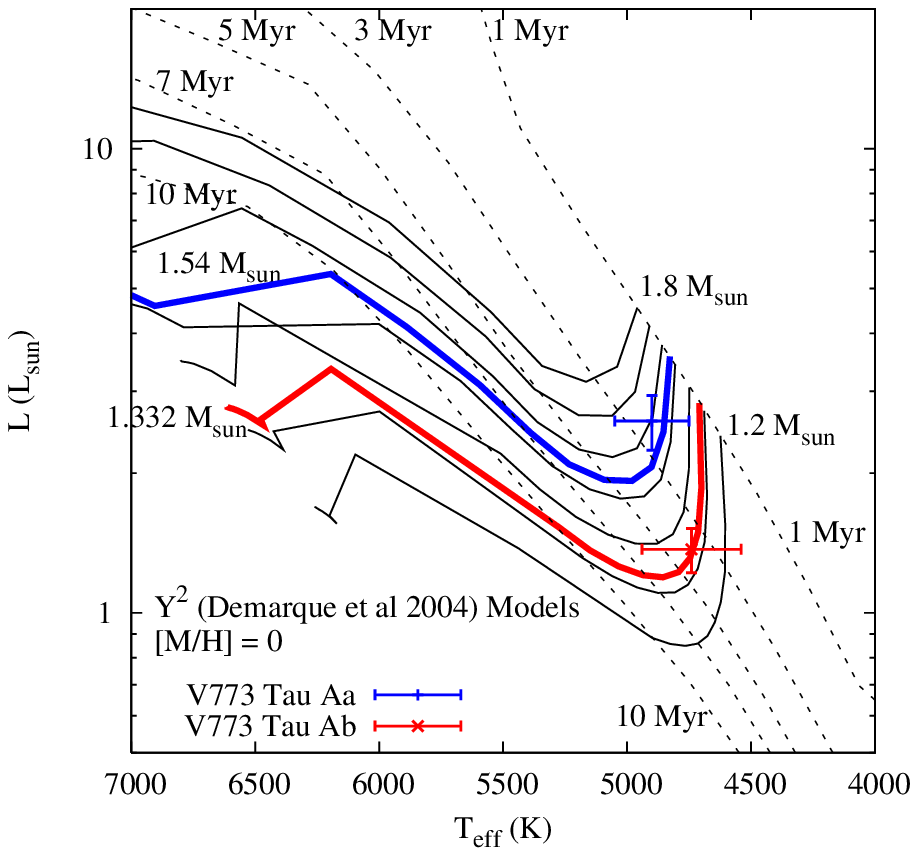}\\
\plottwo{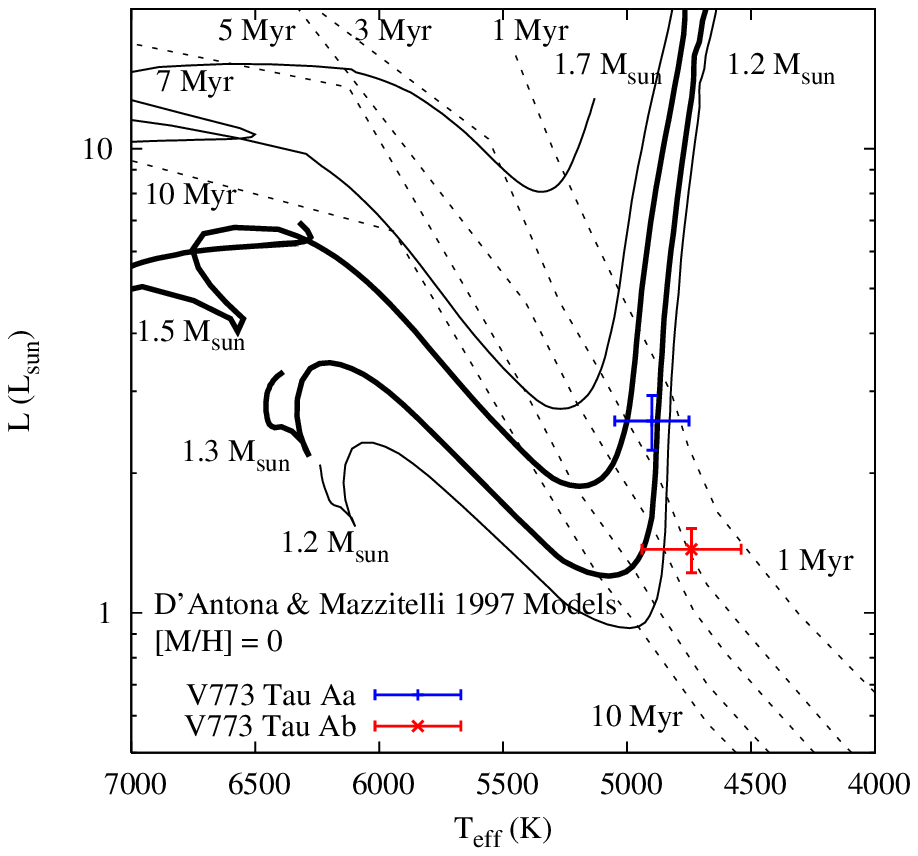}{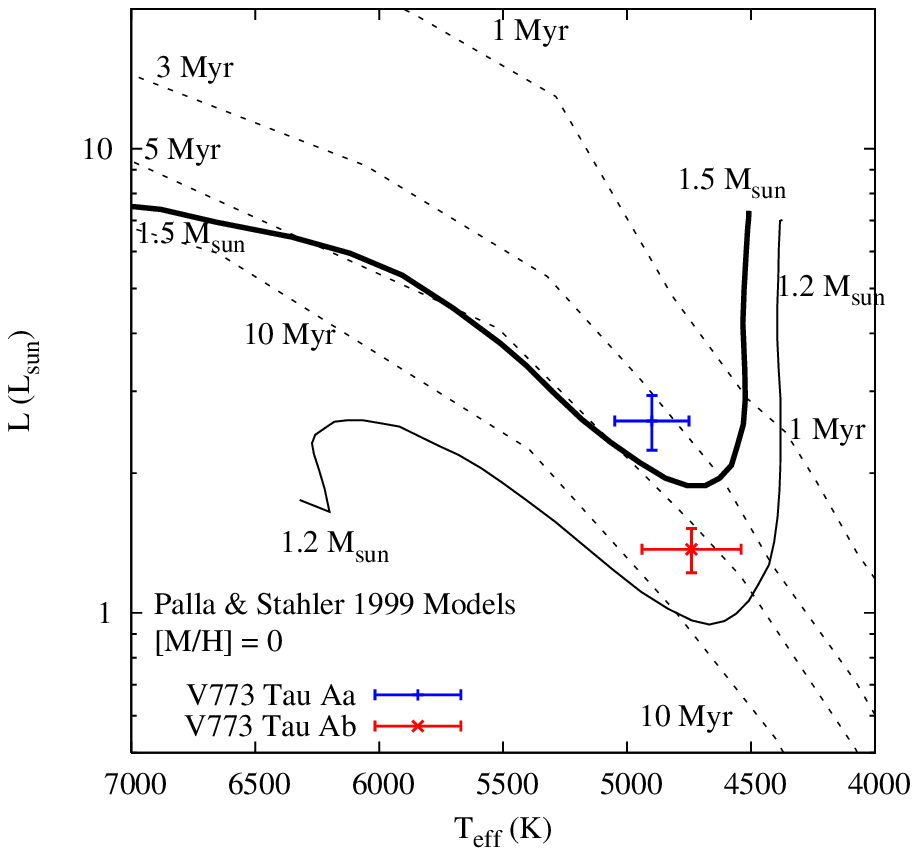}
\caption{V773~Tau~A Components Compared With PMS Models.  Here we show
the V773~Tau~A components in theoretical HR diagrams
(luminosity/$T_{\rm eff}$), and solar-composition PMS evolutionary
tracks and isochrones by \citet{Siess2000} (top left), Y$^2$
\citep{Demarque2004} (top right), \cite{DAntona1997} (bottom left),
and \citet{Palla1999} (bottom right).  All panels show available
tracks in the 1.2 -- 1.8 M$_{\sun}$ range, and isochrones at 1, 3, 5,
7, and 10 Myrs.  In the each panel we highlight the available tracks
that best corresponding to our dynamical mass values; the Y$^2$ models
uniquely allow us to compute tracks at these values.  There is
excellent agreement between our component parameter estimates and the
predictions of the Y$^2$ tracks; other models included here are less
consistent with the inferred component parameters.
\label{fig:modelCompare}}
\end{figure}

Figure~\ref{fig:modelCompare} presents theoretical HR-diagram
(luminosity/$T_{\rm eff}$) comparisons between our determinations of
V773~Tau~A component radiometric properties and solar-composition PMS
models from \citet{Siess2000}, Yonsei-Yale \citep[herein
Y$^2$]{Demarque2004}, \citet{DAntona1997}, and \citet{Palla1999}.
\citep[Popular models by][were excluded only because they do not cover
the mass range for these stars.]{Baraffe1998} These models span a wide
range of physical ingredients and assumptions, some more realistic
than others for the particular mass range of interest.  In these
figures evolutionary tracks at fixed mass between 1.2 and 1.8 M$_\sun$
(in steps of 0.1 M$_\sun$ as available from the relevant models) are
shown in solid lines, with heavy lines indicating tracks at masses
close to dynamical masses from Table~\ref{tab:physics}.  Superimposed
in dotted lines are isochrones at a relevant range of ages (1, 3, 5,
7, and 10 Myrs as available from the relevant models).  Finally, the
Y$^2$ models support computing tracks at user-defined masses; tracks
at the estimated component masses from Table~\ref{tab:physics} are
shown for Aa (1.54 M$_\sun$, blue) and Ab (1.332 M$_\sun$, red).

As expected from \citet{Hillenbrand2004}, all the surveyed PMS models
are in reasonable (e.g. two-sigma) agreement for the V773~Tau~A
component mass range.  The models indicate that the V773~Tau~A
components are still in their quasi-isothermal (``Hayashi'')
contraction phase.  More quantitatively some model predictions seem to
match the A component properties better than others.  For instance,
the solar-abundance \citet{Siess2000} (upper left) and
\citet{Palla1999} (lower-right) tracks near the dynamical masses tend
to under-predict our estimated component temperatures and
luminosities.  In particular for the \citet{Siess2000} models, this is
similar to our previous results for HD~98800~B \citep{Boden2005b}.
This apparent temperature discrepancy seems to reinforce a suggestion
by \citet{Montalban2004} that the solar-abundance \citet{Siess2000}
models appear to be systematically cool compared with other models.
Conversely, the \citet{DAntona1997} tracks (lower-left) at our mass
values appear to slightly over-predict our component temperatures and
luminosity estimates.  The Y$^2$ model predictions (upper-right;
including tracks for the specific masses of the components) are seen
to match our parameter determinations very well.

\begin{figure}[p]
\epsscale{0.7}
\plotone{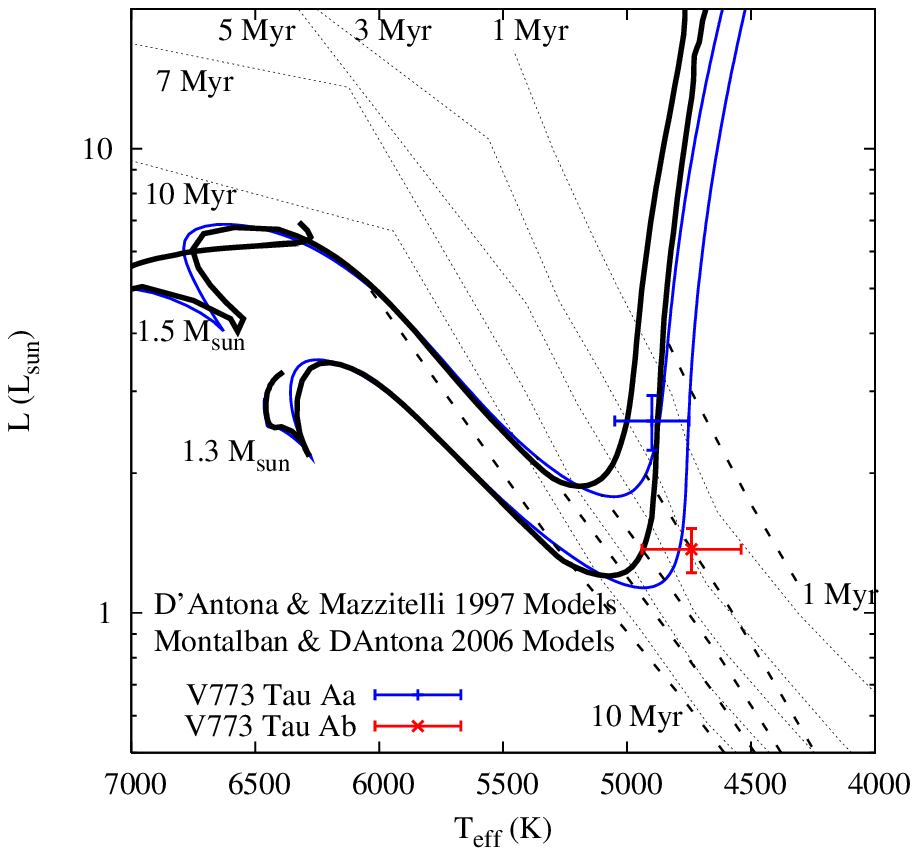}
\caption{V773~Tau~A Components Compared With PMS Models From
\citet{DAntona1997} and \citet{Montalban2006}.  The presentation is
similar to Fig~\ref{fig:modelCompare}, but here we focus on tracks
near the A-component dynamical masses (1.3 and 1.5 M$_\sun$).  The
\citet[][tracks in blue, isochrones in heavy dotted
lines]{Montalban2006} models updated the older \citet[][tracks in
black, isochrones in light dotted lines]{DAntona1997} models with a
revised convection model, bringing the updated models into excellent
agreement with the mass and radiometric parameters estimated here.
\label{fig:modelCompare2}}
\end{figure}

Figure~\ref{fig:modelCompare2} gives a final model comparison.
Recently \citet{Montalban2006} have updated \citet{DAntona1997} models
with a revised ``2D radiative-hydrodynamic'' (2D RHD) convection
treatment.  Figure~\ref{fig:modelCompare2} shows the result of this
updated modeling with tracks from \citet[][in black]{DAntona1997} and
\citet[][in blue]{Montalban2006}; the presentation is similar to that
given in Figure~\ref{fig:modelCompare}, except we have displayed only
mass tracks near the component dynamical mass estimates (1.3 and 1.5
M$_\sun$).  The 2D RHD convective treatment increases the convective
``efficiency'' and energy transport in the atmosphere, moving the
tracks toward lower temperatures during the ``Hayashi'' phase, and
brings the model predictions into excellent agreement with the
component parameters presented here.

Isochrones from the models we considered would predict a range of
component ages from 2 -- 7 Myrs, but restricting attention to the
\citet{Montalban2006} and Y$^2$ models, a more plausible age estimate
is 3 $\pm$ 1 Myrs, with both components appearing coeval to within
present uncertainties.  Such a system age is broadly consistent with
age estimates for the Taurus population using a variety of
evolutionary models \citep[e.g.~WG2001;][]{Palla2002,Luhman2004}.  More
specifically this age estimate would put V773~Tau~A in the putative
intermediate-age population for Taurus as defined by
\citet{Palla2002}.

\subsection{Summary and Conclusions}
\label{sec:conclusions}

We have presented new interferometric (Tables~\ref{tab:V2Table} and
\ref{tab:VLBIastrometry}) and RV data (Table~\ref{tab:RVtable}) on the
PMS binary subsystem V773~Tau~A.  Modeling these data in conjunction
with older VLBI astrometry from P1996 allows us to estimate the visual
and physical orbits for V773~Tau~A, physical parameters of the
components, and place significant constraints on PMS stellar evolution
models.  We characterize the V773~Tau~A orbit model presented here as
{\em preliminary}; we plan additional RV, VLBA, and KI observations
that will further refine the physical orbit model for the A subsystem
and constrain the component parameters.  Further, ongoing RV and
astrometric observations will continue to in order to probe the A-B
orbit, and this in turn will refine the A subsystem RVs and orbit.
Details of A-B orbit modeling will be documented in a future
publication (Torres et al 2007).

The V773~Tau physical parameters inferred from our orbit model are in
good agreement with previous results.  The most notable difference in
our results with previous work is in the system distance; our distance
estimate is at modest variance with the VLBA parallax from L1999.
However, the L1999 data reduction did not account for the binarity of
A which is clearly resolved in multiple epochs (P1996, this work).
With that caveat we believe the L1999 distance and ours are in
acceptable agreement.  Because our distance estimate depends only on
the apparent and physical orbits, we believe it is likely accurate at
its stated uncertainty.  We estimate our relative component mass
errors at 9.1\% and 7.3\% for Aa and Ab respectively, making them
comparable with other PMS stellar dynamical mass determinations
\citep{Hillenbrand2004,Mathieu2007}.

It is noteworthy that the P1996 and present VLBI data agree well with
the A-orbit model presented here (including a model that excludes the
VLBI data themselves, e.g.~see Figure~\ref{fig:VLBAmap}).  Narrowly
this would seem to support the P1996 interpretation that both A
components are active radio emitters.  More broadly, the ability to
compare the radio morphology with at-epoch estimates of the component
positions would seem to present an opportunity to constrain the radio
emission and magnetic field topology in these stars.  A more complete
analysis of the VLBI data in this context will be made in a
forthcoming publication (Massi et al 2007, in prep).

V773~Tau~A radiometric estimates must necessarily account for (and are
complicated by) extinction.  Our estimate of the extinction is in
reasonable agreement with earlier work
\citep[e.g. W1995;][]{Ghez1997,White2001}, but prefers a slightly
higher extinction when constrained by the spectroscopic temperatures
(which are quasi-independent of the extinction).  With this
uncertainty in mind, we have made relatively conservative estimates of
the errors in the radiometric properties.

Even with the relatively conservative error estimates, the combination
of our component dynamical mass and radiometric parameter estimates
appear to differentiate between the range of solar-abundance PMS
models considered here.  In particular, we find better agreement
between the estimated component parameters and predictions of
solar-abundance models from Y$^2$ and \citet{Montalban2006} than other
models considered here, and these two models agree on an V773~Tau age
estimate of approximately 3 $\pm$ 1 Myr; this is in good agreement
with the multiple epochs of Taurus star formation \citep{Palla2002}.
This age would place V773~Tau in the second of three age groups
defined by \citet{Palla2002}, i.e., stars with age $2 < t < 4$ Myr.
Such and younger stars are typically concentrated along the ``central
filaments'' of the Taurus molecular clouds systems.  V773~Tau appears
in fact located at the periphery of a $\sim 0.7\times0.5$ pc-sized
``core'' imaged in the high column density tracing C$^{18}$O molecule
by \citet[their core No.~5]{Onishi1996, Onishi1998}. From the observed
intensity of the C$^{18}$O $J=1-0$ line one can derive an H$_2$ column
density that translates into an A$_V > 5$.  Extinction maps from
Digitized Sky Survey source counts around V773 Tau's position also
indicate an A$_V$ around 5, but have a ``resolution'' of several arc
minutes \citep{Dobashi2005}.  These extinction values are several
times larger than our and other extinction determinations for
V773~Tau~A (\S~\ref{sec:SED}), and seem to indicate that the system is
located just at the near border of a denser cloud core.

Relative to the component parameters estimated here, the
\citet{Siess2000} and \citet{Palla1999} models predict cooler and less
luminous components, while the \citet{DAntona1997} models predict
hotter and more luminous components.  However these conclusions would
change if the component mass estimates were to systematically change
(i.e. both components either more or less massive than present
estimates) by only about two sigma.  Qualitatively this seems
relatively unlikely, but given the uncertainties in this study and the
added complexity of multiple orbits in the V773~Tau system it would be
prudent to withhold final judgment until the modeling of the A-B orbit
improves with time and an expanded astrometric and radial velocity
data set.


Finally, it is important to note that placing constraints on stellar
evolutionary models in general, and PMS evolutionary models in
particular requires precise measurement of more than just component
dynamical mass, but also equally important parameters such as
luminosity, $T_{\rm eff}$, and abundance.  Even at the present level
of the work here (i.e.~dynamical masses better than 10\% 1-sigma),
uncertainties in $L$ and $T_{\rm eff}$ are significant factors in
limiting conclusions we can draw about model predictions.  As for
abundance, we have restricted our comparisons of V773~Tau~A component
properties with solar-composition models.  Fundamentally this is a
practical consideration; some of the model sets considered here only
support solar abundances.  Nominally abundances in Taurus-Auriga (and
by consequence V773~Tau) are thought to be near solar values
\citep[see][]{Padgett1996}.  However, as in other sectors of the HR
diagram, more secure abundance determinations in star forming regions,
along with improved measurements of luminosity and temperature will be
critical for definitive tests of PMS models.

\acknowledgements

Part of this work was performed at the Michelson Science Center (MSC),
California Institute of Technology under contract with the National
Aeronautics and Space Administration (NASA).

Some of the data presented herein were obtained at the W.M.~Keck
Observatory, which is operated as a scientific partnership among the
California Institute of Technology, the University of California and
the NASA.  The Observatory was made possible by the generous financial
support of the W.M.~Keck Foundation.  We gratefully acknowledge the
support of personnel at the Jet Propulsion Laboratory, W.M.~Keck
Observatory, and the MSC in obtaining KI observations of V773~Tau~A.
The authors wish to recognize and acknowledge the very significant
cultural role and reverence that the summit of Mauna Kea has always
had within the indigenous Hawaiian community.  We are most fortunate
to have the opportunity to conduct observations from this mountain.

The CfA RV observations presented here were originally advocated by
Robert Mathieu; we thank him for his gracious contribution of these
data in their use here.  We thank P.~Berlind, M.~Calkins, J.~Caruso,
R.J.~Davis, G.~Esquerdo, J.~Peters, A.~Milone, and R.P.~Stefanik for
obtaining many of the spectroscopic observations.  GT acknowledges
partial support from NASA's MASSIF SIM Key Project (BLF57-04) and NSF
grant AST-0406183.

We thank F.~D'Antona, J.~Montalb\'an, S.~Stahler, and F.~Palla for
sharing their pre-main sequence evolutionary models, and for fruitful
discussions on applying them.  Thanks also to G.~Duchene for sharing
IR photometry for V773~Tau~A (including his kind permission to report
these data here in Table~\ref{tab:Fluxtable}), and the anonymous
referee whose many thoughtful comments helped greatly improve this
manuscript.

This research has made use of services of the MSC at the California
Institute of Technology; the SIMBAD database, operated at CDS,
Strasbourg, France; of NASA's Astrophysics Data System Abstract
Service; and of data products from the Two Micron All Sky Survey,
which is a joint project of the University of Massachusetts and the
Infrared Processing and Analysis Center, funded by NASA and the
National Science Foundation.


\begin{thebibliography}{}



\bibitem[Baraffe et al.(1998)]{Baraffe1998}
Baraffe, I.~et al~1998, \aap~337, 403.


\bibitem[Boden et al.(2000)]{Boden2000}
Boden, A., et al~2000, \apj~536, 880.

\bibitem[Boden et al.(2005a)]{Boden2005a}
Boden, A., Torres, G., \& Hummel, C.~2005a, \apj~627, 464.

\bibitem[Boden et al.(2005b)]{Boden2005b}
Boden, A.~et al~2005b, \apj~635, 442.

\bibitem[Colavita(1999)]{Colavita1999}
 Colavita, M.~1999, \pasp~111, 111.

\bibitem[Colavita et al(2003)]{Colavita2003}
 Colavita, M.~et al~2003, \apj~592, L83.

\bibitem[D'Antona \& Mazzitelli(1997)]{DAntona1997}
D'Antona, F.~\& Mazzitelli, I.~1997, Mem.~S.A.It., 68, 807

\bibitem[Demarque et al(2004)]{Demarque2004}
Demarque, P.~et al~2004, \apjs~155, 667.

\bibitem[Dobashi et al.(2005)]{Dobashi2005}
Dobashi, K.~et al~2005, \pasj, 57, 417.


\bibitem[Duchene et al(2003)]{Duchene2003}
Duchene, G.~et al.~2003, \apj~592, 288 (D2003).


\bibitem[Feigelson et al(1994)]{Feigelson1994}
Feigelson, E.~et al~1994, \apj~432, 373.

\bibitem[Ghez et al(1993)]{Ghez1993}
Ghez, A., Neugebauer, G., \& Matthews, K.~1993, \aj~106, 2005.

\bibitem[Ghez et al(1997)]{Ghez1997}
Ghez, A., White, R., \& Simon, M.~1997, \apj~490, 353.

\bibitem[Hartigan et al(1994)]{Hartigan1994}
Hartigan, P., Strom, K., \& Strom, S.~1994, \apj~427, 961.

\bibitem[Hillenbrand \& White(2004)]{Hillenbrand2004}
Hillenbrand, L.~\& White, R.~2004, \apj~604, 741.



\bibitem[Kurucz(2001)]{KuruczModels} Kurucz, R.~2001, Kurucz models
used here are available at http://cfaku5.cfa.harvard.edu.

\bibitem[Kutner et al(1986)]{Kutner1986}
Kutner, M., Rydgren, A., \& Vrba, F.~1986, \aj~92, 575.


\bibitem[Latham(1992)]{Latham1992}
 Latham, D.\ W. 1992, in IAU Coll.\ 135, Complementary Approaches to
Double and Multiple Star Research, ASP Conf.\ Ser.\ 32, eds.\ H.\ A.\
McAlister \& W.\ I.\ Hartkopf (San Francisco: ASP), 110


\bibitem[Latham et al.(2002)]{Latham:02}
 Latham, D.~W.~et al~2002~\aj~124, 1144.



\bibitem[Leinert et al.(1993)]{Leinert1993}
Leinert, C.~et al~1993, \aap~278, 129.

\bibitem[Lejeune et al(1997)]{Lejeune1997}
 Lejeune, T.~et al~1997, \aaps~125, 229.

\bibitem[Lejeune et al(1998)]{Lejeune1998}
 Lejeune, T.~et al~1998, \aaps~130, 65.

\bibitem[Lestrade et al(1999)]{Lestrade99}
Lestrade, J-F, et al~1999, \aap~344, 1014 (L1999).

\bibitem[Luhman(2004)]{Luhman2004}
Luhman, K.~2004, \apj~617, 1216.

\bibitem[Mason et al(2001)]{Mason2001}
Mason, B.~et al~2001, \aj~122, 3466.

\bibitem[Montalb\'an et al(2004)]{Montalban2004}
Montalb\'an, J.~et al~2004, \aap~416, 1081.

\bibitem[Montalb\'an \& D'Antona(2006)]{Montalban2006}
Montalb\'an, J.~\& D'Antona, F.~2004, \mnras~370, 1823.

\bibitem[Mart\'in et al(1994)]{Martin1994}
Mart\'in, E.~et al~1994, \aap~282, 503.

\bibitem[Massi et al(2002)]{Massi2002}
Massi, M., Menten, M., \& Neidhofer, J.~2002, \aap~382, 152.

\bibitem[Massi et al(2006)]{Massi2006}
Massi, M.~ et al~2006, \aap~453, 959.

\bibitem[Mathieu et al(2007)]{Mathieu2007}
Mathieu, R.~et al~2007, Proc PPV, U.~Ariz Press.

\bibitem[Napier et al(1994)]{Napier1994}
Napier, P.~et al~1994,
Proc IEEE, 82, 658.


\bibitem[Nordstr\"om et al.(1994)]{Nordstrom:94}
 Nordstr\"om, B.~et al~1994, \aap~287, 338.



\bibitem[O'Neal et al(1990)]{O'Neal1990}
O'Neal, D.~et al~1990, \aj~100, 1610.

\bibitem[Onishi et al.(1996)]{Onishi1996}
Onishi, T.~et al~1996, \apj~465, 815.

\bibitem[Onishi et al.(1998)]{Onishi1998}
Onishi, T.~et al~1998, \apj~502, 296.

\bibitem[Padgett(1996)]{Padgett1996}
Padgett, D.~1996, \apj~471, 847.

\bibitem[Palla \& Stahler(1999)]{Palla1999}
 Palla, F.~\& Stahler, S.~1999, \apj~525, 772.

\bibitem[Palla \& Stahler(2001)]{Palla2001}
 Palla, F.~\& Stahler, S.~2001, \apj~553, 299.

\bibitem[Palla \& Stahler(2002)]{Palla2002}
 Palla, F.~\& Stahler, S.~2002, \apj~581, 1194.

\bibitem[Pickles(1998)]{Pickles1998}
Pickles, A.~1998, \pasp~110, 863.

\bibitem[Phillips et al(1996)]{Phillips1996}
Phillips, R.~et al~1996, \aj~111, 918 (P1996).


\bibitem[Rydgren et al(1976)]{Rydgren1976}
Rydgren, A.~Strom, S., \& Strom, K.~1976, \apjs~30, 307.

\bibitem[Siess et al(2000)]{Siess2000}
 Siess L., Dufour E., \& Forestini M. 2000, \aap~358, 593.

\bibitem[Stefanik et al.(1999)]{Stefanik:99}
 Stefanik, R.\ P., Latham, D.\ W., \& Torres, G. 1999, in Precise
Stellar Radial Velocities, IAU Coll.\ 170, ASP Conf.\ Ser., 185, eds.\
J.\ B.\ Hearnshaw \& C.\ D.\ Scarfe (San Francisco: ASP), 354.




\bibitem[Tamazian et al(2002)]{Tamazian2002}
Tamazian, V.~et al.~2002, \apj~578, 925.

\bibitem[Torres et al.(2003)]{Torres2003}
 Torres, G. et al.~2003, \aj~125, 825.

\bibitem[Walker \& Walstencroft(1988)]{Walker1988}
 Walker, H.~\& Walstencroft, R.~1988, \pasp~100, 1509.

\bibitem[Webb et al(1999)]{Webb1999}
 Webb, R.~et al~1999, \apj~512, L63.


\bibitem[Welty(1995)]{Welty1995}
Welty, A.~1995, \aj~110, 776 (W1995).

\bibitem[White \& Ghez(2001)]{White2001}
White, R.~\& Ghez, A.~2001, \apj~556, 265 (WG2001).

\bibitem[Woitas(2003)]{Woitas2003}
Woitas, J.~2003, \aap~406, 685.

\bibitem[Zensus et al.(1995)]{Zensus1995}
Zensus, J., Diamond, P., \& Napier, P.~1995, ASP Conf.~Ser.~ 82: Very
Long Baseline Interferometry and the VLBA

\bibitem[Zucker \& Mazeh(1994)]{Zucker:94}
 Zucker, S.~\& Mazeh, T.~1994, \apj~420, 806.

\bibitem[Zucker et al.(1995)]{Zucker:95}
 Zucker, S., Torres, G.,~\& Mazeh, T.~1995, \apj~452, 863.

\end{thebibliography}
\end{document}